\documentclass[preprint,nofootinbib,
showpacs,preprintnumbers,amsmath,amssymb]{revtex4}
\usepackage{graphicx}

\newcommand{\tr}{\textrm{Tr}}
\newcommand{\be}{\begin{eqnarray}}
\newcommand{\ee}{\end{eqnarray}}
\newcommand{\la}{\langle}
\newcommand{\ra}{\rangle}
\newcommand{\ds}[1]{
#1{\hskip-2.0mm}/
}

\begin{document}
\author{P. Faccioli}
\affiliation{European Centre for Theoretical Studies in 
Nuclear Physics and Related Areas (E.C.T.*),\\
Strada delle Tabarelle 286, Villazzano (Trento) I-38050, Italy.}
\title{A Systematic Study of the Chirality-Mixing Interactions in QCD} 
\vspace {2cm}
\begin{abstract}

We present a study of the QCD interactions 
which do not conserve the chirality of quarks.
These non-perturbative forces are responsible for the 
violation of the $U_A(1)$ charge conservation
and for the breaking of chiral symmetry.
From a systematic analysis we argue that the  
leading sources chirality flips are the interactions mediated 
by topological vacuum field fluctuations.
We study in detail the contribution of instantons and derive 
a simple model-independent semi-classical 
prediction. This result can be used to check
on the lattice 
if instantons are  the dynamical 
mechanism responsible both for chiral symmetry breaking
 and for the anomalous violation of the $U_A(1)$ charge conservation.
\end{abstract}
\maketitle

\section{Introduction and Motivation}

Understanding the structure of the quark-quark interaction at all scales
is a fundamental task of nuclear and high-energy physics, 
which requires the solution of the non-perturbative sector of QCD.
Although this goal is not yet been completely 
achieved, a good deal of information has been gathered in the past decades.
In particular, we known that the non-perturbative 
dynamics mixes quark modes of different chirality.
This property of the strong interaction is signaled by 
the anomalous violation of the $U_A(1)$ charge conservation
and by the breaking of  chiral symmetry.
These two phenomena have
important consequences on the physics of the light hadrons.
The anomalous $U_A(1)$ Breaking (U1B) allows to explain the 
absence of a ninth Goldstone boson.
The Spontaneous Chiral Symmetry Breaking (SCSB)
 shapes the structure of the spectrum 
of the lightest hadrons and their interaction, at low momenta. 
Clearly, identifying the dynamical mechanism responsible for 
the SCSB and the U1B is a fundamental step toward our comprehension
of the quark-quark interaction. 

From the dynamical point of view, the structure of the spectrum of the 
lightest mesons indicates that quark-quark interaction
is particularly attractive in the flavor non-singlet $0^-$ channel and 
less attractive (or possibly, at some distance, even repulsive)
in the singlet $0^-$ channel ($\eta'$).
Very useful insight in the physics of chiral symmetry breaking has come from 
the Nambu Jona-Lasinio (NJL) model \cite{NJL,NJL2}.
In this approach, one postulates an effective quark-quark interaction,
which breaks \emph{spontaneously} the $U_V(3)\times~U_A(3)$ symmetry 
down to $U_V(3)$.  
Such a symmetry breaking pattern would obviously 
generate nine Goldstone bosons, 
being also the $U_A(1)$ symmetry broken spontaneously. 
Hence, in order to reproduce the
observed $\eta-\eta'$ splitting,  an additional 
term is introduced, 
which breaks $U_A(1)$ explicitly and
simulates the contribution from the 't~Hooft determinant interaction. 
Clearly, in such a framework, there are two independent 
sources of violation of the $U_A(1)$ charge, because 
the dynamical origin of the spontaneous $U_A(3)$ breaking and 
of the explicit (i.e. anomalous) U1B are distinct. In fact,
by suppressing the instanton contribution in the Lagrangian, the $\eta-\eta'$ 
splitting  disappears,
yet chiral symmetry remains broken, hence the pions  remain light. 

Conversely, one can imagine the opposite scenario in which \emph{both} 
chiral symmetry breaking and the $\eta-\eta'$ splitting are consequences of the
same dynamical mechanism. 
In other words, one can ask whether the same 
gauge configurations which generate the quark condensate also break 
the $U_A(1)$ symmetry through the axial anomaly. 
In such a scenario, the $U_A(1)$ 
charge conservation is violated only through the  
anomaly. Moreover, there cannot 
be an $\eta-\eta'$ splitting without the SCSB, and vice-versa.

Instantons provide an example of gauge configurations which
solve the $U(1)$ problem \cite{thooftu1}  and, at the same time, 
break chiral symmetry \cite{instantonscsb} (for a review see also 
\cite{dyakonovchiral}). The standing question is whether these 
semi-classical fields  are the  \emph{dominant} configurations in
both phenomena, or if their contribution is only sub-leading.
Unfortunately, since a systematic semi-classical approach to QCD is not 
possible, one has to rely on phenomenological models, such as the 
Instanton Liquid Model (ILM) \cite{shuryak82, shuryakrev}. 
This model was shown 
to quantitatively reproduce the breaking of chiral symmetry in the vacuum,
and its restoration at high temperatures.
Moreover, it also provides a good description of the spectrum 
light hadrons \cite{meson2pt,baryon2pt,baryon2ptint,mymasses} 
and their form factors \cite{blotz,3ptILM,pionFF,nucleonFF}.

On the other hand, a traditional argument against the hypothesis that 
instantons drive both the U1B and the SCSB
is based on the large $N_c$~limit \cite{largenc1}.
In this limit, the topological susceptibility disappears, 
instantons are suppressed by the exponent of the action, 
yet  chiral symmetry remains broken. 
One is then lead to suppose that, if the large $N_{c}$ world is at least
qualitatively similar to the real one, instantons might not be the 
leading mechanism for the SCSB and
that the $\eta-\eta'$ splitting and the SCSB have a different dynamical origin.

In a recent work, Sch\"afer studied in detail the instanton
content of QCD with many colors \cite{schafernin}.
 He observed that the large entropy of 
instantons in $SU(N_c)$ can overcome the exponential suppression
due to the action, in such a way
 that the instanton density can remain finite, at large 
$N_c$.
From an analysis based on numerical simulations and mean-field estimates, he
found that in the ILM the quark condensate is of $O(N_c)$, 
while the $\eta'$ mass is of $O(1/N_c)$. From this facts, he concluded that
 the ILM  is not necessarily in conflict with the large $N_{c}$~analysis.
Although a number of lattice studies seem to confirm a picture in which 
instantons drive the SCSB (see e.g. \cite{chirallatt} and
references therein), a general consensus on these issues
has not yet been reached \cite{draper}, 
and further studies are therefore needed.

In this work, we set-up a framework
to study the time-evolution of the chirality of quarks, 
 and we apply it
to investigate the dynamics underlying the U1B and the SCSB, in QCD.
In the next section, we shall define 
two functions of the Euclidean time,
 $R^{S}(\tau)$ and  $R^{NS}(\tau)$, which measure 
the rate of chirality-flips, in a quark-antiquark system 
with zero total angular momentum, and total isospin 0 and 1, respectively.
We will show that the typical time between two chirality-flipping 
interactions scales with the inverse of the $\eta'$ mass, 
in the $I=0$ channel, and with
the inverse of the mass of the $\delta$ meson (which is the axial partner of 
the pion) in the $I=1$ channel. Moreover, by performing
a spectral analysis of the our probability amplitude ratios, 
we shall show that chirality
flips are dominated by the \emph{anomalous} breaking of $U_A(1)$, 
while the contribution of the SBCS is only sub-leading. 
This result is model independent, because
it is based  only on the numerical values of 
the masses of the lowest lying scalar and pseudo-scalar mesons, which are
known experimentally.
It implies that, in the $|\bar{q}\,q\ra$ system
 under consideration, the leading
chirality mixing interaction is mediated by topologically charged gauge 
fields. 

Then, we will use the Operator Product Expansion (OPE)
to perform a systematic study of the dynamical origin of the 
mixing of chirality in QCD.
We shall see that, if vacuum field fluctuations were small 
(vacuum dominance approximation), then the 
chirality flips would be dominated by
the SCSB, and the contribution from the anomalous U1B would 
be completely negligible. 
However, this prediction is not supported by 
phenomenology which, as we have mentioned above, seems to favor 
the opposite scenario in which the anomalous U1B dominates.
Therefore, one is lead to conclude that
vacuum fluctuations play a major role in the mixing of chirality.
Collecting all these observations, we shall argue that 
the leading dynamical source of helicity flips is represented by 
the interaction mediated by some large topological vacuum fluctuations, which
badly violate the factorization assumption.

In section \ref{instantons}  we will address the question of whether such
dynamics is driven by instantons.
We shall first compute analytically the leading instanton contribution to the 
helicity-flip probability amplitudes, in the ILM.
Then, we shall derive a simple semi-classical prediction, 
which does not depend on the phenomenological parameters of the ILM.
Such a relationship is a model-independent signature of the instanton-induced
interaction and can be used to check in an unambiguous way
if instantons are responsible both for the SCSB and for the U1B.
The feasibility of checking such a signature using 
lattice simulations will be discussed. 

Interestingly,  we find that the single-instanton
effects generate 
the same rate of chirality flips, in the singlet and non-singlet 
channel.
The equality $R^{S}(\tau)=R^{NS}(\tau)$ 
(which holds for small values of  $\tau$, for which many-instanton 
effects can be neglected) is non-trivial and represents  
another model independent signature of instanton-induced forces.

Results and conclusions of this work are summarized in 
section \ref{conclusions}. 

\section{Mixing of quark chirality in QCD}
\label{chiralflip}

For sake of simplicity, let us consider QCD with two flavors. Moreover, 
since we are interested in the helicity flips generated by the quark-quark 
interaction, it is convenient to work in the chiral limit. This way, the purely
 kinematical chirality flips induced by the quark mass are suppressed.

We begin by defining the following combination of gauge 
invariant correlation functions:
\be
\label{AflipNS}
A^{NS}_{flip}(\tau) &:=& \la 0|\,T[\,\bar{u}(\tau)\,P_L\,d(\tau)\, 
  \bar{d}(0)\,P_R\,u(0)\,| 0 \ra + (P_L \leftrightarrow P_R)\nonumber\\
&+& \la 0|\,T[\,\bar{d}(\tau)\,P_L\,u(\tau)\, 
  \bar{u}(0)\,P_R\,d(0)\,| 0 \ra + (P_L \leftrightarrow P_R)
\ee
\be
\label{ASflip}
A^{S}_{flip}(\tau)&:=& \frac{1}{2}
[\la 0|\,T[\,\bar{u}(\tau)\,P_L \,u(\tau)\, 
 \bar{d}(0)\,P_R\,d(0)\,| 0 \ra + (P_R \leftrightarrow P_L)
\nonumber\\
&+& \la 0|\,T[\,d(\tau)\,P_L\,d(\tau)\, 
\  \bar{d}(0)\,P_R\,d(0)\,| 0 \ra.+ (P_R \leftrightarrow P_L)
\nonumber\\
&+& \la 0|\,T[\,\bar{d}(\tau)\,P_L\,d(\tau)\, 
 \bar{u}(0)\,P_R\,u(0)\,| 0 \ra + (P_R \leftrightarrow P_L)
\nonumber\\
&+& \la 0|\,T[\,u(\tau)\,P_L\,u(\tau)\, 
   \bar{u}(0)\,P_R\,u(0)\,| 0 \ra.+ (P_R \leftrightarrow P_L)]
\ee
where,
\begin{eqnarray}
P_R:=\frac{1+\gamma_5}{2}\qquad
P_L:=\frac{1-\gamma_5}{2}.
\end{eqnarray}

$A^{NS}_{flip}(\tau)$ and $A^{S}_{flip}(\tau)$ denote
 the probability amplitude for a flavor
singlet (non-singlet) $|q~\bar{q}\ra$ state 
to be found, after a time interval $\tau$, in a state in which the 
chirality of the quark and antiquark is 
flipped\footnote{In all formulas, analytic continuation to Euclidean time
is assumed.}. 
In QCD, even in the case in which $m_u=m_d=0$, 
quark states with different chirality can mix under time evolution.
Hence, we expect such matrix elements to be in non-vanishing, in general.

Similarly, we define:
\begin{eqnarray}
\label{AnonflipNS}
A^{NS}_{non-flip}(\tau) &:=& \la 0|\,T[\,\bar{u}(\tau)\,P_L\,d(\tau)\, 
  \bar{d}(0)\,P_L\,u(0)\,| 0 \ra + (P_L \leftrightarrow P_R)\nonumber\\
&+& \la 0|\,T[\,\bar{d}(\tau)\,P_L\,u(\tau)\, 
  \bar{u}(0)\,P_L\,d(0)\,| 0 \ra + (P_L \leftrightarrow P_R)
\ee
\be
\label{AnonflipS}
A^{S}_{non-flip}(\tau)&:=& \frac{1}{2}
[\la 0|\,T[\,\bar{u}(\tau)\,P_L \,u(\tau)\, 
 \bar{d}(0)\,P_L\,d(0)\,| 0 \ra + (P_L \leftrightarrow P_R)
\nonumber\\
&+& \la 0|\,T[\,d(\tau)\,P_L\,d(\tau)\, 
  \bar{d}(0)\,P_L\,d(0)\,| 0 \ra.+ (P_L \leftrightarrow P_R)
\nonumber\\
&+& \la 0|\,T[\,\bar{d}(\tau)\,P_L\,d(\tau)\, 
 \bar{u}(0)\,P_L\,u(0)\,| 0 \ra + (P_L \leftrightarrow P_R)
\nonumber\\
&+& \la 0|\,T[\,u(\tau)\,P_L\,u(\tau)\, 
   \bar{u}(0)\,P_L\,u(0)\,| 0 \ra.+ (P_L \leftrightarrow P_R)],
\end{eqnarray}
These functions measure the probability amplitude
for the quark and antiquark not to have exchanged their chirality, 
after a time interval $\tau$.

\begin{figure}
\includegraphics[scale=0.3,clip=]{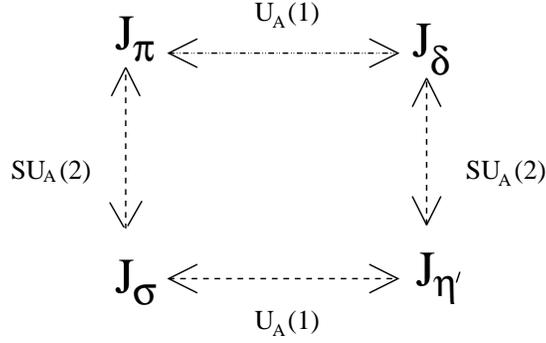}
\caption{$U_A(3)$ transformations relating the 
$J_\pi,\,J_\delta,\,J_{\eta'}$, and $J_\sigma$ operators.}
\label{u3}
\end{figure}

It is convenient to rewrite such matrix elements in terms of scalar and 
pseudo-scalar mesonic two-point correlation functions:
\begin{eqnarray} 
A^{NS}_{flip}(\tau) &=& \Pi_{\pi}(\tau)-\Pi_{\delta}(\tau)\\
A^{NS}_{non-flip}(\tau) &=& \Pi_{\delta}(\tau)+\Pi_{\pi}(\tau),
\ee
\be
A^{S}_{flip}(\tau) &=& \Pi_\sigma(\tau)-\Pi_{\eta'}(\tau)\\
A^{S}_{non-flip}(\tau) &=& \Pi_\sigma(\tau)+\Pi_{\eta'}(\tau),
\end{eqnarray}
where the correlation functions are defined
 as\footnote{In this expression, have denoted with 
$\eta'$ an iso-singlet pseudo-scalar state, i.e.
the $SU(N_f=2)$ correspondent of the $|\eta_0\ra$ state. 
Moreover we can, without
loss of generality, choose $\tau> 0$ 
and disregard the Euclidean time-ordering.}:
\begin{eqnarray}
\label{PIpi}
\Pi_\pi(\tau)     &=&\la 0 |J_{\pi}(\tau)
\,J_{\pi}^{\dagger}(0)| 0 \ra\\ 
\label{PIeta}
\Pi_{\eta'}(\tau)&=&\la 0 |J_{\eta'}(\tau)\,J_{\eta'}^{\dagger}(0)| 0 \ra\\ 
\label{PIdelta}
\Pi_{\delta}(\tau)&=&\la 0 |J_{\delta}(\tau)\,J_{\delta}^{\dagger}(0)|\, 
0 \ra\\
\label{PIsigma}
\Pi_\sigma(\tau)&=&\la 0 |J_{\sigma}(\tau)\,J_{\sigma}^{\dagger}(0)| 0 \ra. 
\end{eqnarray}
The interpolating operators, exciting states with given $(I,J^p)$ quantum
numbers, are defined~as:
\begin{eqnarray}
J_{\pi}(\tau)&:=&\bar{u}(\tau)\, i\,\gamma_5 d(\tau)\\
\label{Jpi}
J_{\delta}(\tau)&:=&\bar{u}(\tau)\,d(\tau)\\
\label{Jdelta}
J_{\eta'}(\tau)&:=& \frac{1}{\sqrt{2}}\,
\left(\bar{u}(\tau)\,i\,\gamma_5 u(\tau)+\bar{d}(\tau)
\,i\,\gamma_5 d(\tau)\right)\\
\label{Jeta}
J_{\sigma}(\tau)&:=&\frac{1}{\sqrt{2}}\,
\left(\bar{u}(\tau)\, u(\tau)+\bar{d}(\tau)\,d(\tau)\right)
\label{Jsigma}
\end{eqnarray}
At this point it is worth observing that the operators
 $J_\pi,\,J_\delta,\,J_{\eta'}$, and $J_\sigma$  can be transformed
into each other  by means of appropriate $U_A(3)$ transformation (see Fig. 
\ref{u3}).

With these amplitudes, we can now construct two ratios 
which represent the probability amplitude to find the quarks in 
the flipped chirality state, 
relative to the amplitude for remaining in the same state.
In the singlet and non-singlet channels, we have:
\begin{eqnarray}
\label{RS}
R^{S}(\tau)=\frac{A^{S}_{flip}(\tau)}
  {A^{S}_{non-flip}(\tau)}=\frac{\Pi_\sigma(\tau)-\Pi_{\eta'}(\tau)}
{\Pi_{\sigma}(\tau)+\Pi_{\eta'}(\tau)}\\
\label{RNS}
R^{NS}(\tau)=\frac{A^{NS}_{flip}(\tau)}
  {A^{NS}_{non-flip}(\tau)}=\frac{\Pi_{\pi}(\tau)-\Pi_\delta(\tau)}
{\Pi_{{\delta}}(\tau)+\Pi_\pi(\tau)}
\end{eqnarray}
In the chiral limit, these ratios  carry information 
about the rate of chirality mixing interactions, in QCD.

Let us study (\ref{RS}) and (\ref{RNS}) in the limit of large 
Euclidean times. In this case, 
one expects quarks to interact several times.
Intuitively, there should exist a characteristic time scale $\tilde{\tau}$, 
determining how often chirality 
mixing interactions occur in the QCD vacuum.
After having exchanged their chirality many times, quarks essentially 
``loose the memory''  of what was their initial state, therefore they 
should be found in either chirality configuration, with equal 
probability. That is to say:
\be
R^{NS}(\tau), R^{S}(\tau)\stackrel{\tau\gg\tilde{\tau}}{\rightarrow} 1.
\ee

In order to derive this result rigorously and to determine $\tilde{\tau}$, 
we use the fact that, 
at large enough $\tau$, the spectral representation  of the
correlation functions
 (\ref{PIpi}-\ref{PIsigma}) is dominated by the contribution of
the lowest lying states
with the appropriate quantum numbers \cite{shu}:
\be
\Pi_\pi(\tau)&\stackrel{(\tau\to\infty)}{\rightarrow}&
\,\lambda_\pi^2\,\frac{m_\pi}{4\,\pi^2\,\tau}\,
K_1(m_\pi\,\tau)\\
\Pi_{\delta}(\tau)&\stackrel{(\tau\to\infty)}{\rightarrow}&
\lambda_{\delta}^2\,\frac{m_{\delta}}{4\,\pi^2\,\tau}\,
K_1(m_{\delta}\,\tau)\\
\label{PIsigmaspec}
\Pi_\sigma(\tau)&\stackrel{(\tau\to \infty)}{\rightarrow}&
2\, |\la 0 |\bar{q} q | 0 \ra|^2+\,\lambda_\sigma^2\,\frac{m_\sigma}
{4\,\pi^2\,\tau}\,
K_1(m_\sigma\,\tau),\\
\Pi_{\eta'}(\tau)&\stackrel{(\tau\to \infty)}{\rightarrow}&
 \lambda_{\eta'}^2\frac{m_{\eta'}}{4\,\pi^2\,\tau}\,K_1(m_{\eta'}\,\tau)
\ee
where $\lambda_{\pi,{\delta},\sigma\,\eta'}$ 
are the coupling constants of the densities (\ref{Jpi}-\ref{Jsigma}) 
to the corresponding lowest lying states.

Hence, in the large $\tau$ limit,
we can rewrite the ratio $R^{NS}(\tau)$ in the following way:
\be
\label{Rxi}
R^{NS}(\tau)=
\frac{1-\xi(\tau)}{1+\xi(\tau)}
\stackrel{(\tau\to\infty)}{\rightarrow}1-2\,\xi(\tau)+
 o(\xi^2),
\ee
where
\be
\label{xi}
\xi(\tau):= \frac{\Pi_{\delta}(\tau)}{\Pi_\pi(\tau)}\stackrel{(\tau\to 
\infty)}{\rightarrow}
\left(\frac{\lambda_{\delta}}{\lambda_\pi}\right)^2\,
\sqrt{\frac{m_{\delta}}{m_\pi}}
e^{-(m_{\delta}-m_\pi)\,\tau},
\ee

These relationships show that the information about the 
initial chirality of the quarks is exponentially 
destroyed, as the time increases.
From the exponent in (\ref{xi}) 
can immediately read-off the corresponding characteristic time scale:  
\be
\tilde{\tau}^{NS}=\frac{1}{(m_{\delta}-m_\pi)}\sim~0.2 ~\textrm{fm}.
\label{tauNS}
\ee

We shall now prove that the ratio $R^{NS}(\tau)$ represents the amplitude for 
the chirality-flips induced 
by the non-conservation of the $U_A(1)$ charge. 
If the $U_A(1)$ symmetry  is not spoiled, then we have
$m_\pi=m_\delta$ and $\lambda_\pi=\lambda_\delta$.
As a consequence $\xi(t)\to~1$ and 
the chirality-flip amplitude ratio (\ref{RNS})
tends to zero, at infinity.
This condition 
 is sufficient to imply that $R^{NS}(\tau)$ is identically 
zero.
In fact, let us suppose that this function is not
identically zero, yet vanishes at infinity. This corresponds to the
unphysical scenario in which quarks 
can flip their chirality only in some finite time interval, 
but eventually have to ``go back'' to their initial chirality state.

The fact that $\Pi_\pi(\tau) \ne \Pi_\delta(\tau)$ and therefore 
$R^{NS}(\tau)\ne0$ only implies that the axial
charge is not conserved, regardless if this is predominantly 
because of a spontaneous or anomalous breaking of the $U_A(1)$ 
symmetry\footnote{
Indeed, in the NJL model, in which the $U_A(1)$ is broken spontaneously,
one has $\Pi_\pi(\tau)>\Pi_\delta(\tau)$. 
We thank T.Sch\"afer for pointing this out.}.
The combination of correlators that relates directly to the anomaly is the 
difference between the $\pi$ and $\eta'$ two-point functions,
$\Pi_\pi(\tau)-\Pi_{\eta'}(\tau)$.
It is therefore convenient to add and subtract the 
$\Pi_{\eta'}(\tau)$ correlator,
from the numerator in $R^{NS}(\tau)$:
\be
R^{NS}(\tau)=\frac{\Pi_\pi(\tau)-\Pi_{\eta'}(\tau)}
{\Pi_{\pi}(\tau)+\Pi_\delta(\tau)}+
\frac{\Pi_{\eta'}(\tau)-\Pi_{\delta}(\tau)}{\Pi_{\pi}(\tau)+\Pi_\delta(\tau)}
=:R^{NS}_{anom.}(\tau) + R^{NS}_{SU_A}(\tau).
\ee
We are now in condition to disentangle the violation of the axial 
charge conservation induced by  the anomalous and by the 
spontaneous breaking of $U_A(1)$.
Indeed, if there was no anomaly, and the $U_A(1)$ symmetry was just 
spontaneously broken, then $\Pi_\pi(\tau)=\Pi_{\eta'}(\tau)$ and 
 $R^{NS}_{anom.}(\tau)$ would vanish. 
In this case, the contribution to $R^{NS}(\tau)$ would come entirely 
from $R^{NS}_{SU_A}(\tau)$. 
Conversely, if the $SU_A(N_f)$ symmetry was not broken, then we would have
$\Pi_{\eta'}(\tau)=\Pi_{\delta}(\tau)$ (see Fig. \ref{u3}),
therefore $R_{SU_A}(\tau)=0$.
In the real world we observe a  relatively large $\pi-\eta'$ 
splitting. As a consequence, $R^{NS}(\tau)$ turns out to be dominated
by the anomalous term,  $R^{NS}_{anom.}(\tau)$.
To see this, we use again the 
spectral representation, in the large Euclidean time limit:
\be
\label{RA}
R^{NS}_{anom.}(\tau)&=&~1-C_{\pi\,\eta'}\,e^{-(m_\eta'-m_\pi)~\tau} -
C_{\pi\,\delta}\,e^{-(m_\delta-m_\pi)~\tau}+...,\\
\label{RB}
R^{NS}_{SU_A}(\tau)&=&~C_{\eta'\,\pi}\,e^{-(m_\eta'-m_\pi)~\tau}
\left(1-C_{\delta\,\pi}~e^{-(m_\delta-m_\pi)~\tau}
-C_{\eta'\,\pi}~e^{-(m_{\eta'}-m_\pi)~\tau}\right)+...,
\ee
where
\be
C_{h\,h'}:=\left(\frac{\lambda_h}{\lambda_{h'}}\right)^2
\sqrt{\frac{m_h}{m_h'}},
\ee
and ellipses denote higher order terms in $e^{-(m_\delta-m_\pi)~\tau}$.
We can see that, in this limit, $R^{NS}_{anom.}(\tau)$ is of order
1, while $R^{NS}_{SU_A}(\tau)$ is exponentially suppressed by 
the $\eta'-\pi$ mass difference.
We conclude that  the violation of the axial charge conservation, parametrized
by the function $R^{NS}(\tau)$,
is mainly due to the \emph{anomalous} breaking of the $U_A(1)$ symmetry.
This conclusion has quite important implications
on quarks and gluons dynamics. It suggests that, in this channel,
 the leading chirality mixing interaction is mediated by 
topologically charged gauge fields.

For sake of completeness, let us now consider the helicity-flip 
amplitude ratio in the singlet channel:
\be
\label{Repsilon}
R^{S}(\tau)=
\frac{1-\epsilon(\tau)}{1+\epsilon(\tau)}
\stackrel{(\tau\to\infty)}{\rightarrow}1-2\,
\epsilon(\tau)+ o(\epsilon^2),
\ee
where,
\be
\label{epsilon}
\epsilon(\tau):= \frac{\Pi_{\eta'}(\tau)}{\Pi_\sigma(\tau)}
\stackrel{(\tau\to \infty)}{\rightarrow}
\frac{\lambda_{\eta'}^2\, \sqrt{m_{\eta'}}}{\sqrt{2}\,8
\,\la 0 |\bar{q}{q}| 0 \ra^2}\,\frac{e^{-m_{\eta'}\,\tau}}{(\pi\,\tau)^{3/2}}.
\ee

In this channel, we could not find a simple way to single-out the 
effects due to the anomalous U1B. 
This is essentially because the scalar density operator (\ref{PIsigma}) 
has a non-vanishing vacuum expectation value 
(Eq. (\ref{PIsigmaspec})).
As a consequence, $R^{S}(\tau)$ receives an explicit
contribution from the quark 
condensate\footnote{It is tempting to simply replace $\Pi_\sigma(\tau)$
with a modified scalar two-point functions, in 
which the vacuum contribution has been subtracted out, 
$\Pi'_\sigma(\tau):= \Pi_\sigma(\tau)-2\la0|
\bar{q}\,q|0\ra^2$. However, the corresponding amplitude ratio,
$R'^{S}(\tau):= (\Pi'_\sigma-\Pi_{\eta'})/(\Pi'_\sigma + \Pi_{\eta'})$
would not have a simple probabilistic 
interpretation in terms of chirality flips.}(Eq. (\ref{epsilon})). 
Notice that the anomalous U1B contribution induce an exponential mixing of
chirality, while the SCSB participates only through an additional pre-exponent.
The  characteristic 
time, which  determines the exponential mixing of chirality is the inverse
of the $\eta'$ mass:
\be
\label{taus}
\tilde{\tau}^S= \frac{1}{m_{\eta'}}.
\ee
Notice that $\tilde{\tau}^S,\tilde{\tau}^{NS}\ll
\frac{1}{\Lambda_{QCD}}$, which  
suggests that the non-perturbative physics associated with the U1B and 
SCSB is characterized by an additional scale, significantly
larger than $\Lambda_{QCD}$.

The phenomenological analysis performed so far has indicated 
that the helicity flips in the non-singlet channel are
dominated by the dynamics responsible for the anomalous U1B. 
Now we want investigate what can be learnt from this result about the
non-perturbative quark-quark dynamics.  At this purpose, we need to consider 
$R^{NS}(\tau)$ in the opposite
limit of small Euclidean times,  where quark and gluons are the relevant
degrees of freedom.
After performing Wick contractions, the correlation functions
 (\ref{PIpi}-\ref{PIsigma}) read:
\be
\label{PIpiwick}
\Pi_\pi(\tau)&=&\la \tr~[ S(\tau,0)\,\gamma_5\,S(0,\tau)\,\gamma_5 ]\ra,\\
\label{PIdeltawick}
\Pi_\delta(\tau)&=&- \la \tr~[ S(\tau,0)\,S(0,\tau) ]\ra,\\
\label{PIetawick}
\Pi_{\eta'}(\tau)&=&~[\la \tr~[ S(\tau,0)\,\gamma_5\,S(0,\tau)\,\gamma_5 ]\ra 
- 2~\la\tr~[ S(\tau,\tau)\,\gamma_5]\,
\tr[ S(0,0)\,\gamma_5]\ra],\\
\label{PIsigmawick}
\Pi_\sigma(\tau)&=&[2~ \la~\tr~[ S(\tau,\tau)]\,\tr[ S(0,0)]\ra 
-\la \tr~[ S(\tau,0)\,S(0,\tau)]\ra],
\ee
where $S(\tau,0)$ is the quark propagator and 
$\la \cdot\ra$ denotes the average over all gauge configurations. 
At extremely small distances, asymptotic freedom demands that 
the correlators (\ref{PIpi}-\ref{PIsigma}) approach those evaluated in the free
theory. One has:
\begin{equation}
\label{PIfree}
\Pi_\pi(\tau), \Pi_{\eta'}(\tau), \Pi_{\delta}(\tau), 
\Pi_\sigma(\tau)\stackrel{(\tau\to 0)}{\to} \Pi_0(\tau):=
\frac{N_c}{\pi^4\, \tau^6}.
\end{equation}
Consequently, we have 
$R^{S}(\tau),R^{NS}(\tau)\stackrel{(\tau\to 0)}{\to}0 $. This is 
an expected result: in a free theory of massless quarks, chirality
is a conserved quantum number, and the rate of quark helicity flips is zero. 

As $\tau$ gradually increases from zero, perturbative corrections to 
the correlators (\ref{PIpi}-\ref{PIsigma}) become more and more important. 
However,  their contribution to (\ref{RS}) and (\ref{RNS})
has to vanish exactly, because the  
perturbative quark-gluon vertex does not mix quark modes with opposite
chirality. 
Hence, any deviation from zero of the ratios (\ref{RS}) and (\ref{RNS})
is a signature of non-perturbative physics.

Non-perturbative QCD corrections, in the limit $\tau\to 0$, 
can be evaluated in a systematic way, 
using the Operator Product Expansion (OPE). It  known  
 that, in the
chiral limit, the quark and gluon  
condensate singular terms in OPE cannot distinguish 
between scalar and pseudo-scalar mesonic correlators \cite{SVZ,scalarSVZ}.
Hence, including the contribution up to dimension 4 operators
one finds $R^{NS}(\tau)=0$.

In the 't~Hooft picture, in which the anomaly is realized by instantons,
the fact that such singular terms cannot reproduce the 
chirality flipping due to the
non-conservation of the axial charge is an expected result. 
In fact, the leading instanton contributions to the correlators 
(\ref{PIpi}-\ref{PIdelta}) tend to a constant, in the $\tau\to0$ limit.  
Therefore, one expects some deviation from zero in $R^{NS}(\tau)$ to appear,
starting from the regular terms, i.e. from dimension 6 operators.
Indeed, we find\footnote{In this expression and in (\ref{Delta})
we have used isospin symmetry and we have 
neglected logarithmic perturbative corrections. Notice that the
term proportional to the quark condensate and that proportional 
to the mixed quark condensate
$g\la0|\bar{q}G_{\mu\,\nu}\sigma_{\mu\,\nu}q|0\ra$ are suppressed, in the chiral limit.}:
%
\be
R^{NS}_{anom.}(\tau)\simeq \frac{-2~\pi^4}{N_c}\la0| \bar{u}\,i\gamma_5\,u
\,\bar{d}\,
i\gamma_5\,d |0\ra \, \tau^6 +...\\
R^{NS}_{SU_A}(\tau)\simeq  
\frac{\pi^4}{N_c}~\left[~\left(\frac{\mbox{}}{\mbox{}}
\la0| \bar{u}\,i\gamma_5\,u\,
\bar{u}\,i\gamma_5\, u |0\ra~-~\la0| \bar{u}\,u\,\bar{u}\,u |0\ra\right)\right.
\nonumber\\
\left.
+\left(\frac{\mbox{}}{\mbox{}}\la0| \bar{u}\,i\gamma_5\,u\,
\bar{d}\,i\gamma_5\, d |0\ra~+~\la0| \bar{u}\,u\,\bar{d}\,d |0\ra\right)
\right]\tau^6+...,
\ee

Let us now discuss the implications of this result.
Quite interestingly, we have found that the quantity:
\be
\label{Delta}
\Delta:&=&\lim_{\tau\to 0 }~\left|
\frac{R^{NS}_{SU_A}(\tau)}{R^{NS}_{anom.}(\tau)}\right|=\nonumber\\
&=&\left|\frac{\la0| \bar{u}\,i\gamma_5\,u\,
\bar{u}\,i\gamma_5\, u |0\ra + \la0| \bar{u}\,i\gamma_5\,u\,
\bar{d}\,i\gamma_5\, d |0\ra + \la0| \bar{u}\,u\,\bar{d}\,d |0\ra - 
\la0| \bar{u}\,u\,\bar{u}\,u |0\ra}{2~\la0| \bar{u}\,i\gamma_5\,u
\,\bar{d}\,
i\gamma_5\,d |0\ra}\right|
\ee
parametrizes the relative contribution to quark helicity flips 
of the spontaneous symmetry breaking with
respect to the anomalous symmetry breaking. In particular, if
 $\Delta\simeq~0$, then 
the violation of the $U_A(1)$ charge comes almost entirely from the anomaly.

It is instructive to study the behavior of the chirality flip  ratio
$R^{NS}(\tau)$, in the large $N_c$ limit. 
In this limit, the spectral representation of the 
condensates is dominated by vacuum insertions and one can 
use the factorization approximation:
\be
\la 0|\bar{q}~\Gamma_1 q\,\bar{q}~\Gamma_2\,q|0\ra\simeq
\frac{1}{{\mathcal N}^2}(
\tr[\Gamma_1]\,\tr[\Gamma_2]-\tr[\Gamma_1\,\Gamma_2])
\,\la0|\bar{q}\,q|0\ra^2,\qquad {\mathcal N}:=4\,N_f\,N_c,
\ee
which gives $\la0| \bar{u}\,i\gamma_5\,u
\,\bar{d}\,i\gamma_5\,d |0\ra\simeq 0$, hence $\Delta\to \infty$. 
This result implies that, in the asymptotically large $N_c$ world, 
the chirality flips are dominated by the spontaneous chiral symmetry breaking,
while the anomalous contribution disappears, in agreement with common wisdom.

In the real world, the accuracy of the factorization approximation is 
questionable. Indeed, the analysis performed at large distances indicates that
the spin flips are actually dominated by the anomaly and not by the SCSB 
(Eqs. (\ref{RA}) and (\ref{RB})). 
In general, the vacuum dominance assumption is expected to
fail in in the presence of large vacuum field fluctuations.
Hence, our analysis suggests that in QCD the helicity flips are mainly induced
by some large vacuum gauge field fluctuations.
As we mentioned above, such fluctuations have to be topologically charged, 
as they induce the U1B though the anomaly. 
The most natural candidates are instantons. Their contribution will be 
analyzed in the next section.

\section{Instanton-induced chirality flips}
\label{instantons}

In the previous section we have argued that 
the helicity flips are dominated by some large field fluctuations which
have connection with the axial anomaly.
Instantons are example of gauge fields which, at the same time,
provide a realization of the anomaly and break chiral symmetry. 
It is therefore interesting to estimate their contribution to the ratio
(\ref{RNS}).
At short distances, the leading instanton contribution to the 
relevant correlation
functions (\ref{PIpi}-\ref{PIdelta})
can be computed analytically, using the Single Instanton Approximation 
(SIA).
This is an effective theory of the instanton vacuum (for a detailed description
of this approach, see \cite{SIA}), in which the contribution
of the closest instanton is taken into account explicitly, while all other
(infra-red) multi-instanton degrees of 
freedom are integrated out and replaced by a single effective
parameter $m^*$, defined as:
\be
\label{mstar}
m^*=
\left[\left(\int\,d\rho\, d(\rho)
\frac{\bar{n}}{5\, \pi^2\,\rho^4}
\right)
\frac{1}{
\left<
\left[
\textrm{Tr}
\sum_{I,J} \psi_{0\,I}(x)
\left(
\frac{1}{T}
\right)_{I\,J}
\psi_{0\, J}^\dagger(x)
\right]^2
\right>}\right]^{1/2},
\ee
where $d(\rho)$ is the instanton size distribution, $\bar{n}$
 is the instanton density,  $\psi_{0\,I(x)}$ is the zero mode wave function in 
the field of the $I$-th pseudo-particle and
$T_{I\,J}$ is the overlap matrix, given by:
\be
T_{I J}=\int d^4 z \psi^\dagger_{0\,I}(z)(i\ds{\partial})\psi_{0\,J}(z).
\ee
The effective parameter $m^*$, computed numerically 
in the ILM from  (\ref{mstar}), is of the order of 70~MeV.
The main advantage of the SIA is that it allows to obtain the leading instanton
contributions to Green's functions by means of simple analytical calculations.
The range of applicability of this approach
was studied in detail \cite{SIA,3ptILM}.
It was found to be very accurate for correlation
functions smaller than the typical distance between two neighbor instantons.

The evaluation of the correlation functions (\ref{PIpi}-\ref{PIdelta}) leads to
\footnote{For sake of simplicity, in this calculation we have replaced
 the non-zero mode part of the quark propagator with the free 
propagator (zero-mode approximation). It is possible to show that this
 approximation 
is very accurate for the particular correlation functions we are
 considering \cite{mythesis}. 
The main results of this section (Eq. (\ref{sia1}) and 
Eq.(\ref{directI}) below)
 do not change, when one correctly accounts for the non zero-mode 
contributions.}:
\be
\Pi^{SIA}_{\pi}&=&\Pi_0(\tau) + \Pi_{OI}(\tau)\\
\Pi^{SIA}_{\delta}&=&\Pi_0(\tau) - \Pi_{OI}(\tau)\\
\ee
where $\Pi_0(\tau)$ is the usual free 
 correlator defined in (\ref{PIfree})
and $\Pi_{OI}(\tau)$ is the instanton contribution, given by:
\be
\label{PIOI}
\Pi_{OI}(x)=\frac{4\,\bar{n}}{m^{*\,2}\,\pi^4}\,
\int d^4 z~\int\,d\,\rho~ d(\rho)~\frac{\rho^4}{[z^2+\bar{\rho}^2]^3\,
((z-x)^2+\bar{\rho}^2)^3}, 
\ee
The corresponding result for the chirality-flip amplitude ratio is:
\be
\label{equality}
R_{SIA}^{NS}(\tau)= \frac{\Pi_{OI}(\tau)}{\Pi_0(\tau)},
\ee

In passing, we observe that the leading
instanton contribution  to $R^{S}(\tau)$ 
is the same as that to $R^{NS}(\tau)$:
\be
\label{sameR}
R_{SIA}^{S}(\tau)=R_{SIA}^{NS}(\tau).
\ee
This equality, which is valid only in the region in which the
one-instanton effects are dominant, is non-trivial. 
In fact, we recall that at large distances 
these quantities have very different spectral representations. 
Hence, in general, we expect them to be different functions of $\tau$.

Let us now establish the connection with the OPE analysis of the previous
section. Expanding Eq. (\ref{PIOI}) for short Euclidean times 
one finds:
\be
\label{sia1}
R^{NS}_{SIA}(\tau)=
\frac{\pi^2\,\bar{n}}{15\,m^{*\,2}}\,
\int\,d\rho \, d(\rho)\,\frac{1}{\rho^4}\,\tau^{6}+...
\ee
where the ellipsis denote terms which are higher orders in $\tau$.

\begin{figure}
\includegraphics[scale=0.4,clip=]{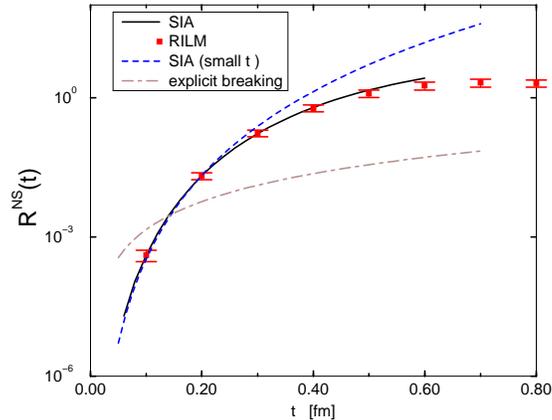}
\caption{The  
chirality flipping amplitude ratio, in the non-singlet
channel, $R^{NS}(\tau)$. Points are numerical simulations in the RILM, the 
solid line denotes the SIA prediction (\ref{equality}), and the dashed line 
its short-time expression (\ref{sia1}).The dot-dashed line 
represents the explicit breaking contribution (\ref{explicit}), with
$m=150$~MeV.}
\label{RatioR}
\end{figure}

In general, calculations of instantonic effects require the knowledge of 
two quantities which cannot be obtained in a systematic way in QCD: 
the instanton density $\bar{n}$, 
and the instanton size distribution 
$d(\rho)$\footnote{We recall that the effective parameter $m^*$ 
can be obtained 
from $d(\rho)$ and $\bar{n}$  through Eq. 
(\ref{mstar}).}.  
From a phenomenological estimate, Shuryak suggested to use\cite{shuryak82} 
$\bar{n}\simeq~1~\textrm{fm}^{-4}$,  
and $d(\rho)=\delta(\rho-1/3\textrm{fm})$.
In Fig. (\ref{RatioR}) the SIA prediction
(\ref{equality}) for $R^{NS}(\tau)$
is compared with its short-time expansion  (\ref{sia1}), 
and with the result of numerical simulations in the Random 
Instanton Liquid Model (RILM)\footnote{For a detailed description of this 
model and a comparison with other versions of the ILM,
 see \cite{shuryakrev}.}, in which  many-instantons 
effects are taken explicitly into account.
We observe that the very simple SIA analytic prediction agrees with the 
much more complicated RILM calculation up to quite large times $\tau\simeq\,
0.6~\textrm{fm}$. 
Moreover, we observe that the short-time expression (\ref{sia1})
converges up to distances of the order of 
$\tau\lesssim~0.25~\textrm{fm}$. This
provides an estimate of the radius of convergence of OPE expansions, 
when the direct-instanton term is included.

The results discussed so far depend on the phenomenological parameters
of the ILM.
Now we make one more step and show that it is possible 
to circumvent all the model dependence\footnote{As another example of 
model-independent instanton prediction
in QCD is discussed in \cite{mymasses}. In that paper, we computed
the proton and pion 
dispersion curves and we derived, in a parameter-free way, 
the instanton contribution to the nucleon mass.}, and obtain the 
semi-classical prediction.
At this purpose, we observe that the instanton contribution to the 
four-quark condensate  reads:
\be
\label{fourquark}
\la0|\bar{q}\,q\,\bar{q}\,q|0\ra_{SIA}=
\frac{2~\bar{n}}{5\,\pi^2\,m^{*\,2}}\,
\int\,d\rho\,d(\rho)\frac{1}{\rho^4}=
\la0|\bar{q}\,\gamma_5~q\,\bar{q}\,\gamma_5\,q|0\ra_{SIA}
\ee
We notice that  leading instanton effects contribute equally 
to the pseudo-scalar and scalar four-condensates and therefore badly
violate the factorization assumption. This implies that, in the
instanton vacuum,  $\Delta\simeq0$. 
Therefore,  the instanton-induced U1B is channeled exclusively through the
axial anomaly, in agreement with the results of the phenomenological 
analysis presented in the previous section.

Combining (\ref{fourquark}) with (\ref{sia1}) we can include
all unknown quantities appearing in our calculation in the expression for the
quark condensate. This way, we obtain
the direct-instanton contribution to OPE for (\ref{RS}) and (\ref{RNS}):
\be
\label{directI}
R_{D.I.}^{NS}(\tau)=
\frac{\pi^4}{6}\,\la0| q\,\bar{q}\,q\,
\bar{q}|0\ra \,\tau^{6}+ ...,
\ee
It is worth emphasizing that this semi-classical calculation
requires no assumption on the 
instanton size and density, yet predicts a particular 
relationship between different vacuum expectation values.
Hence, if the semi-classical contribution is large, so that 
instantons are the leading configurations
responsible for the U1B and the SCSB, 
then this relationship 
must be at least approximatively satisfied, 
also when the quantum average is performed over 
\emph{all} configurations (and not only over the semi-classical 
fluctuations).

Lattice QCD represents the natural framework in which performing such
 a test. On the practical level, however, 
up-to-date simulations still need to use quite
large values of the current quark masses. 
Far from the chiral limit,
the explicit breaking of chiral symmetry provides a competing source of
chirality flips, which has to taken into account. 
The importance of this effect can be estimated by evaluating 
(\ref{RNS}) in the free theory - using a massive quark 
propagator - and then comparing with the ILM prediction (\ref{sia1}).
After an elementary calculation, we obtain:
\be
\label{explicit}
R_{explicit}^{NS}(\tau)=\frac{m^2}{4}\,\tau^2+..., 
\ee
where $m$ is the current quark mass. Clearly, at short distances, the 
contribution to helicity flips
coming from the explicit breaking of chiral symmetry (\ref{explicit}) 
will necessarily dominate over the dynamical effects discussed 
above. 

The key question is if there exists a window of Euclidean times in which
the instanton effects are dominant over the
kinematic ones, and the single-instanton prediction  (\ref{directI}) is 
reliable. 
In Fig. (\ref{RatioR}) the contribution of 
explicit chiral symmetry breaking
 (\ref{explicit}) with a current
mass $m=150$~MeV (of the order of those used in a typical 
lattice calculation) is 
compared with the SIA and RILM estimates.
We conclude that, 
in the region $0.1~\textrm{fm}\lesssim~\tau\lesssim~0.25~\textrm{fm}$,
the instanton contribution should be dominant. 
Moreover, in this window, the relevant correlation functions
(\ref{PIpi}-\ref{PIdelta}) are of order 1, so it 
should not be numerically very 
challenging to measure this signal on the lattice.

\section{Conclusions}
\label{conclusions}
In this paper we have presented a study of the mixing of 
chirality under time evolution, in QCD. 
In the chiral limit, quark  helicity-flips have a purely dynamical origin,
therefore they carry information about the quark-quark interaction.

We have constructed two combinations of correlation functions, $R^{NS}(\tau)$ 
and $R^{S}(\tau)$,
which are related to the  rate of
chirality flips in a quark-antiquark pair, propagating in the QCD vacuum.
Such functions receive no contribution from any number 
of perturbative gluon exchanges, therefore they 
represent  useful tools to investigate the non-perturbative dynamics.

We have shown that the chirality flipping processes
 contributing to $R^{NS}(\tau)$
are induced by the same non-perturbative forces
responsible for the non-conservation of the axial charge. 
In fact, the ratio $R^{NS}(\tau)$ 
would be identically zero, if the degeneracy between the pion 
and its $U_A(1)$ partner was not lifted. 
Using the spectral decomposition at large Euclidean times, 
we have studied the relative contribution to the U1B coming from the 
axial anomaly and from the possible spontaneous breaking of $U_A(1)$.
We found that the U1B
comes almost entirely from the axial anomaly.
From a such a spectral analysis,  we  have also 
obtained an estimate of the characteristic
time between two consecutive spin-flipping 
interactions in QCD and found 
$\bar{\tau}\sim~0.2~\textrm{fm}$.
This value suggests that the
chirality mixing dynamics is characterized by a
non-perturbative length scale which is significantly smaller than the typical 
confinement scale, $1/\Lambda_{QCD}\simeq~1~\textrm{fm}$.

We have calculated the first non-perturbative corrections to the
amplitude ratio $R^{NS}(\tau)$, using OPE.
We have found that these are associated
to the regular terms, which are commonly
related to the so-called ``direct instanton'' contribution.
This is consistent with the 't~Hooft solution of the $U(1)$ problem, in which
the axial anomaly is realized by instantons.
We observed that the fact that $R^{NS}(\tau)$ is dominated by
the anomaly implies that vacuum field fluctuations are large.
In fact, in the vacuum dominance approximation, in which 
fluctuations are neglected, the chirality flips are dominated
by the SCSB, in disagreement with what we found phenomenologically
from the spectral analysis. 
This suggests that the helicity flips are induced by 
topological gauge field fluctuations.

We  calculated analytically 
the instanton contribution to $R^{S}(\tau)$ and
 $R^{NS}(\tau)$ in the ILM, using the SIA.
Then, we derived a model-independent semi-classical 
relation, which connects scalar and 
pseudo-scalar correlators to the four-quark condensates.
This prediction represents a clean signature of instanton-induced dynamics.
We suggest that it should be checked with an \emph{ab initio} calculation, 
on the lattice.
We discussed the feasibility of performing such a calculation
and concluded that the artifacts related to large quark mass effects
should be negligible.

An application of the same framework to the study the
quenching effects in lattice QCD is in preparation.

\acknowledgements

I would like to thank T.Sch\"afer for reading the manuscript and making 
very useful remarks.
It is also a pleasure to acknowledge interesting discussions with
 L. Girlanda, J. Negele, G. Ripka, S. Simula, E. Shuryak,
M. Traini and W. Weise.
{}
\end{document}